\documentclass[draft]{article}
\usepackage{latexsym,zj2008}

\usepackage{amssymb,amsfonts,amscd,amsthm,amsmath}       

\usepackage[utf8]{inputenc}     
\usepackage[english]{babel}     

\usepackage[perpage,symbol*]{footmisc}

\usepackage[final]{graphicx}     

\usepackage{cite}

\usepackage{colortbl}
\usepackage{color}

\def\fr#1#2{\textstyle\frac{#1}{#2}}
\def\ve{\varepsilon} \def\eps{{\epsilon}}
\def\be{\begin{equation}}            \def\ee{\end{equation}}
\def\ba#1{\begin{array}{#1}}         \def\ea{\end{array}}

   \def\k{\kappa}
\def\cL{{\cal L}} \def\l{\lambda}  \def\L{\Lambda}

\begin{document}

\renewcommand{\topmargin}{1.1in}
\renewcommand{\headsep}{0.25in}
\renewcommand{\topskip}{10pt}
\renewcommand{\footskip}{0.35in}

\newcommand{\rum}{\rule{0.5pt}{0pt}}
\newcommand{\rub}{\rule{1pt}{0pt}}
\newcommand{\rim}{\rule{0.3pt}{0pt}}
\newcommand{\numtimes}{\rub\mbox{\raisebox{1.5pt}{${\scriptscriptstyle \times}$}}}
\newcommand{\Boldsq}{\vbox{\hrule height 0.7pt
\hbox{\vrule width 0.7pt \phantom{\footnotesize T}%
\vrule width 0.7pt}\hrule height 0.7pt}}

\newcommand{\tms}{\fontencoding{OT1}\fontfamily{txr}\fontseries{m}\fontsize{10pt}{12pt}\selectfont}
\newcommand{\palsans}{\fontencoding{OT1}\fontfamily{pxss}\fontseries{m}\fontsize{10pt}{12pt}\selectfont}

\renewcommand{\refname}{\rule{50mm}{0.4pt}}
\renewcommand{\tablename}{\small Table}
\renewcommand{\figurename}{\small Fig.}
\renewcommand{\contentsname}{Contents}


\begin{center}
\renewcommand{\baselinestretch}{0.93}
{\Large\bfseries On relevance of modified gravities
 }\par
\renewcommand{\baselinestretch}{1.0}
\vspace*{14pt} Ivan L.\ Zhogin\footnote{Institute  of Solid State
 Chemistry, 18 Kutateladze, 630128 Novosibirsk, Russia.
E-mail: zhogin@mail.ru} \par \vspace*{16pt}
{\footnotesize\parbox{9cm}{%
\textbf{Abstract:}\quad
 It is shown that modified gravity theories with a Lagrangian
composed of the three quadratic invariants of the Riemann
curvature tensor are not appropriate. The field equations are
either incompatible and/or irregular [like $f(R)$-gravities], or,
if compatible,
 lead to the linear instability of polarizations relating to the
 Weyl tensor.
  A more relevant modification is the frame field theory,
 the best and unique
 variant of Absolute Parallelism; it has no free parameters
 ($D\,{=}\,5$ is a must)
 and no singularities arising in solutions. I sketch few
 remarkable features of this theory.
 }}
\medskip
\end{center}

\label{zj-2011-01}

\thispagestyle{empty}

\setcounter{section}{0}
\setcounter{equation}{0}
\setcounter{figure}{0}
\setcounter{table}{0}
\setcounter{page}{231}

\markboth{Ivan L.\ Zhogin}{\thepage}
 \markright{Ivan L.\ Zhogin}

\noindent {\bfseries\S1.\hspace{0.7em}Riemann-squared modified
gravities.}\quad
  To my thinking, all Riemann-squared gravities with a Lagrangian
 composed of the three invariants
quadratic in the Riemannian curvature are not appropriate [as well
as $f(R)$-gravities, $f(R)\,{\neq}\, R$].

 So,
$(a\,{+}\,bR\,{+}R^2\!/\!2)$-gravity ($R$ is the Ricci scalar)
leads to an incompatible  system  of equations (the trace part,
$\mathbf{E}_\mu{}^\mu\,{=}\,0$, can be used  that to remove the
term with $R_{;\l;\l}$ --- excepting the case $D\,{=}\,1$):
\[
\mathbf{E}_{\mu\nu}=R_{;\mu;\nu} - R_{\mu\nu}(b+R)+g_{\mu\nu}
 (a/ 2 + bR/2+ R^2\!/4 -R_{;\l;\l})=0,
\]
\[
\mathbf{E}^\star_{\mu\nu}=\mathbf{E}_{\mu\nu}
 + \varkappa\, g_{\mu\nu} \mathbf{E}_{\l}{}^{\l}=
 R_{;\mu;\nu} -
R_{\mu\nu}(b+R)+g_{\mu\nu}f^\star\!(R)=0;
\]
some people prefer to move the principal derivatives (4-th order)
to the right-hand side (RHS), perhaps trying to hide them in the
energy-momentum tensor; however it is not good way to deal with
 partial differential equations (PDEs).
  The next combination of prolonged equations,
\[ \mathbf{E}^\star_{\mu\nu;\lambda} -
 \mathbf{E}^\star_{\mu\lambda;\nu } =0, \]
after cancellation of the principal derivatives (5-th order),
gives
  new 3-d order equations which are irregular in the second jets:
 the term
 \[ R_{;\ve}R^{\ve}{}_{\mu\nu\lambda} \]
 can not be cancelled by the other terms which contain only
 the Ricci tensor
 and scalar.
The rank of the new subsystem depends on the second derivatives,
 $g_{\mu\nu,\lambda\rho}$
(for the definition of PDE regularity  see \cite{pommaret}).

 As a rule, researchers of modified gravities concentrate their
 efforts on the most symmetrical problems like cosmological
 solutions of spherical symmetry. In this case, the absence of
 regularity
 of the above system is safely masked: the new subsystem becomes just
 an identity due to the skew symmetry of its two indices.

Also irregular in second jets are equations of Gauss-Bonnet (or
Lovelock) gravity with [the] extra dimension(s).

The most interesting case, $R_{\mu\nu} G^{\mu\nu}$-gravity (the
Ricci tensor is contracted with the Einstein tensor), gives the
following compatible system:
 \be \label{rg}  -\mathbf{D}_{\mu\nu}=
 G_{\mu\nu;\lambda}{}^{;\lambda}+
 G^{\epsilon \tau} (2R_{\epsilon
\mu\tau\nu } - \fr12g_{\mu\nu}R_{\epsilon \tau }) =0;
 \ \, \mathbf{D}_{\mu\nu;\lambda}g^{\nu\lambda}\equiv 0\,.
\ee
 In linear approximation, there are simple evolution
equations for the Ricci tensor and scalar
 [there are $D(D\,{-}\,1)/2\,$
polarizations in total --- including one scalar polarization]:
\[ \square R=0, \ \square R_{\mu\nu}=0. \]
Using the Bianchi identity,
 $R_{\mu\nu[\lambda\epsilon;\tau]}\,{\equiv}\, 0$,
  its prolongation and contractions,
\[ R_{\mu\nu[\lambda\epsilon;\tau];\rho} g^{\tau\rho}\equiv0, \
 R_{\mu\nu[\lambda\epsilon;\tau]}g^{\mu\tau}\equiv0, \]
one can write the evolution equation
 for the Riemann (or Weyl) tensor
(the linear approximation again):
 \be \label{riem} \square
 R_{\mu\epsilon\nu\tau}=R_{\mu\nu,\epsilon\tau}-
 R_{\mu\tau,\nu\epsilon}+R_{\epsilon\tau,\mu\nu}
 -R_{\epsilon\nu,\mu\tau}.
 \ee
 This equation is more complex: it has the source term (in its
 RHS)
 composed from the Ricci tensor. As a result, in the general case when
 the Ricci-polarizations do not vanish, the polarizations
 relating to the Weyl tensor [and responsible for gravitational
 waves and tidal
 forces; their number is $D(D\,{-}\,3)/2$] should grow linearly with
 time,
 \[a(t) = (c_0 + c_1 t)\exp(- i \omega t),\]
  while the linear
 approximation is valid.

This means that the regime of weak gravity is linearly unstable,
as well as the trivial solution itself
 (\emph{i.e.}, in this theory,  \emph{nothing}
 is not so \emph{real}).
  Hence,
this theory is physically irrelevant, because we are still living
in conditions of very weak gravity. (Note that in General
Relativity, when the Ricci tensor is expressed
 through  the energy-momentum tensor
 [which does not expand into plane waves ---
 with the dispersion law of light in a
 vacuum],
   equation~(\ref{riem}) describes the process of
  gravitational wave generation.)

The linear instability here does not contradict the correctness of
the Cauchy problem; the modern compatibility theory (see the
Pommaret's book \cite{pommaret}) gives easy answers about the
Cauchy problem, number of polarizations, and so on (especially
easy for analytical PDE systems).

And the last remark. Let $\cL\,{=}\,\sqrt{-g}\,L$ is a homogeneous
Lagrangian density of degree $p$ (as a function of the metric,
$g_{\mu\nu}$); that is,
\[  \cL(\k g_{\mu\nu})=\k^p \cL(g_{\mu\nu})\, . \]
The result of variation, the symmetric tensor
 \[ \mathbf{D}^{\mu\nu} =
 \frac{\delta \cL}{\sqrt{-g}\,\delta g_{\mu\nu}}, \
 \mathbf{D}^{\mu\nu}{}_{;\nu}\equiv0, \]
 has the next relation to its Lagrangian (the trace is
 proportional to the Lagrangian scalar up to a covariant divergence):
 \be\label{lagr} \sqrt{-g}\,\mathbf{D}^{\mu\nu} g_{\mu\nu}
 \equiv p\, \cL + (\sqrt{-g}\, A^{\mu})_{,\mu}, \ \mbox{ or \ }
\mathbf{D}^{\mu\nu} g_{\mu\nu} \equiv
  p\, L + A^{\mu}{}_{;\mu}
  \ee

If $ \mathbf{D}_{\mu\nu}$  (with
$\mathbf{D}^{\mu\nu}{}_{;\nu}\,{\equiv}\,0$) is found just by
using the Bianchi identity, this relation reveals the
corresponding Lagrangian, see {\em{e.g.}}\
 equation (\ref{rg}).

\bigskip\noindent
{\bfseries\S2.\hspace{0.7em}Frame field theory.
 Co- and contra-singularities
  in AP and the unique equation.}\quad
 The theory of frame field, $h^a{}_\mu$, also known as Absolute
Parallelism (AP), has a large symmetry group which includes both
the global symmetries of Special Relativity (defining the
signature) and the local symmetries,  pseudogroup {\it Diff}$(D)$,
of General Relativity.

 AP is more appropriate as a
modified gravity, or just a good theory with topological charges
and quasi-charges (their phenomenology, at some conditions and to
the certain extent, can look
 like a quantum field theory) \cite{on,tc}. In this
case,  the Ricci tensor has very specific form (due to field
equations; linear approximation):  
\[ R_{\mu\nu}\propto \Phi_{(\mu,\nu)}, \, \
 \
 \Phi_\mu=h_a{}^\nu (h^a{}_{\nu,\mu}-h^a{}_{\mu,\nu})
\ \mbox{is the torsion's trace};
\]
this form does not cause the Weyl polarizations growth, see (1);
so the  weak gravity regime (but not the trivial
solution!$\cdot\!$) is stable.

  There is one unique equation of AP (non-Lagrangian, with a
unique $D$) which solutions are free of arising singularities.
 The formal integrability (compatibility) test
 \cite{pommaret} can be extended to the cases of degeneration of
either co-frame matrix, $h^a{}_\mu$ (co-singularities),  or
contra-variant frame (or frame density of some weight), serving as
the local and covariant test for singularities. This test singles
out the next equation (and $D\,{=}\,5$  \cite{on};
$\eta_{ab}\,{=}\,\mbox{diag}(-1,1,\ldots,1)$, then
 $h\,{=}\,\det
h^a_{\ \mu}\,{=}\,\sqrt{-g}$):
 \begin{equation} \label{ue}
 {\bf E}_{a\mu}=L_{a\mu\nu;\nu}- \fr13 (f_{a\mu}
 +L_{a\mu \nu }\Phi _{\nu })=0\, ;
\end{equation}
 here \qquad \qquad  \qquad $ L_{a\mu \nu }=L_{a[\mu \nu]}=
\Lambda_{a\mu \nu }-S_{a\mu \nu }-\fr23 h_{a[\mu }\Phi_{\nu]},$
 \[
\Lambda_{a\mu \nu }=2h_{a[\mu;\nu]}, \ S_{\mu \nu \l}=3\L_{[\mu
\nu \l]}, \ \Phi_\mu=\L_{aa \mu}, \ f_{\mu\nu}=2\Phi_{[\mu;\nu]}=
2\Phi_{[\mu,\nu]} .
\]
 Coma ``,'' and semicolon ``;'' denote partial derivative and
covariant differentiation with symmetric Levi-Civita connection,
respectively. One should retain the identities (for further
details see \cite{on,tc}):
 \be\label{iden}
 \Lambda_{a[\mu\nu;\lambda]} \equiv 0\,,
  \ \  h_{a\l}\Lambda_{abc;\l}\equiv f_{cb}\,
  (= f_{\mu\nu}h_c^{\,\mu} h_b^{\,\nu}), \ f_{[\mu\nu;\l]}\equiv0
  .\ee
 Equation ${\bf E}_{a\mu;\mu}=0$ gives
 `Maxwell-like equation' (sometimes, in covariant expressions,
 I omit
 $\eta_{ab}$ and
 $g^{\mu\nu}\,{=}\,h_a^{~\mu} h_b^{\ \nu} \eta^{ab}$
 in contractions):
 \[
(f_{a\mu}
 +L_{a\mu \nu }\Phi _{\nu })_{;\mu}=0,\]
\begin{equation}\label{max}{
 \mbox{ or \ }
 f_{\mu\nu;\nu}=(S_{\mu \nu\l }\Phi _{\l })_{;\nu} \ \
[= -\fr1 2 S_{\mu \nu\l }f_{\nu\l},
 \mbox{ see eq-n (\ref{skew}) below} ] \, .}
\end{equation}
In reality, eq-n (\ref{max}) follows from the symmetric part only,
because the skewsymmetric one gives an identity; note also that
the trace part
 becomes irregular   if $D\,{=}\,4$ (forbidden $D$;
 the principal derivatives vanish):
 \be\label{skew}
2{\bf E}_{[\nu\mu]}=S_{\mu\nu\l;\l}=0, \ {\bf
E}_{[\nu\mu];\nu}\equiv 0; \ee
 \be\label{trace} {\bf E}_{\mu\mu}={\bf
E}_{a\mu}h_b^\mu\eta^{ab} =\fr{4-D}3 \Phi_{\mu;\mu}
 -\fr12\L_{abc}^2+\fr13S_{abc}^2+\fr{D-1}9\Phi_a^2=0.
  \ee
System (\ref{ue}) remains compatible under adding $f_{\mu\nu}=0$,
see (\ref{max});  this is not the case for the other covariants,
$S, \Phi$, or the Riemann curvature; the last relates to tensor
$\L$ as usually:
 \[ R_{a\mu\nu\lambda}= 2h_{a\mu;[\nu;\lambda]}; \
h_{a\mu}h_{a\nu;\l}=\fr12 S_{\mu\nu\l}-\L_{\l\mu\nu}.\]

GR is a special case of AP. Using 3-minors
 ({\em i.e.}, corank-3 minors) of co-metric,
  \[  [\mu \nu ,\varepsilon \tau ,\alpha \beta ] \equiv
 \partial^3 (-g)/(\partial g_{\mu \nu }
\partial g_{\varepsilon \tau }\partial g_{\alpha \beta }) ,\,\]
and their skew-symmetry features, one can write the vacuum GR
equation
   as
follows:  $\, 2(-g)G^{\mu \nu }\,{=}\,[\mu \nu ,\varepsilon \tau
]_{,\ve\tau}\,{+} (g'^2\!)\,{=}$
 \be \label{gr}{
 =
 [\mu \nu ,\varepsilon \tau , \alpha
\beta ](g_{\alpha \beta ,\varepsilon \tau }+ g^{\rho \phi }\Gamma
_{\rho ,\varepsilon \tau } \Gamma _{\phi ,\alpha \beta })=
 \fr12[\mu \nu ,\varepsilon \tau , \alpha
\beta ]R_{\alpha \varepsilon \tau  \beta} =0.}\ee
 Similarly,  all AP equations  can be reshaped
 that 2-minors of co-frame,
\[ \Bigl(\!\!%
\begin{array}{c}
  \mu \,\, \nu \\   a \,\, b\\
\end{array}%
\!\!\Bigr) =\frac{\partial^2 h}{\partial h^{a}_{\ \mu}
\partial h^{b}_{\ \nu}}
= 2!\, h h^{\,\ \mu}_{[a} h^{\,\nu}_{b]}, \]
  \[ \mbox{that is, \ } \
 [\mu_1 \nu_1 ,\ldots, \mu_k \nu_k  ]=\frac1{k!}
 \Bigl(\!\!%
\begin{array}{c}\mu_1\,\cdots\, \mu_k \\
 a_1\,\cdots \, a_k \\ \end{array}%
\!\!\Bigr)
 \Bigl(\!\!%
\begin{array}{c}\nu_1\,\cdots\,\nu_k\\
  a_1 \, \cdots \, a_k \\  \end{array}%
\!\!\Bigr),
 \] completely define the coefficients at the principal
derivatives.

 For example, the simple compatible equation (see Einstein-Mayer
classification of compatible equations in 4$D$ AP \cite{eima}),
 \be \label{estar}  {\bf
E}^*_{a\mu}=\L_{a\mu\nu;\nu}=0, \ {\bf E}^*_{a\mu;\mu}\equiv0 \ee
 gives \cite{on}
 \[{ h^{2}{\bf E}^*_{a}{}^{\mu }=-gg^{\alpha
\mu}g^{\beta \nu} (h_{a\alpha ,\beta \nu }-h_{a\beta,\alpha  \nu }
)+\cdots 
 = h_{a\alpha ,\beta \nu }[\vspace{1mm}\alpha
\mu ,\beta \nu \vspace{-1mm}] +(h^\prime{}^{2})\ .}\]
 Like the determinant,
$k$-minors ($k\,{\leq}\, D$)
 are multi-linear expressions in
 elements of co-frame matrix, $h^a{}_\mu$, and some minors do not
vanish when rank$\,h^a{}_\mu\,{=}\,D{-}1$.

 For any AP equation [including
 eq-ns~(\ref{gr}) and (\ref{estar})],
 with the \emph{unique exception}, eq-n~(\ref{ue}),
  (where only the skew-symmetric
 part participates in the identity and can be written
 with 2- and 3-minors,
 while the symmetric part needs 1-minors which vanish too
 simultaneously with decreasing of co-frame rank), the
  principal terms keep regularity  (and
 the symbol $G_2$ remains involutive \cite{on})
  if\,
 rank$\,h^a_{\ \mu}\,{=}\,D{-}1$.

This observation is important and relevant to the problem of
singularities; it means seemingly that the unique equation
(\ref{ue}) does not suffer of nascent co-singularities in
solutions of general position.

The other case is contra-singularities \cite{on} relating to
degeneration of the contra-variant density of some weight:
 \be
\label{dens} { H_a{}^\mu= h^{1/D_*} h_a{}^\mu;
 H=\det H^a{}_\mu, \
h_a{}^\mu= H^{1/(D-D_*)} H_a{}^\mu\, .} \ee
 Here $D_*$ depends on the choice of
equation: $D_*=2$ for GR, $D_*=\infty$ for eq-n~(\ref{estar}), and
$D_*=4$ for the unique equation (which can be written 3-linearly
in $H_a{}^\mu$ and its derivatives \cite{on,tc}).  %

If integer, $D_*$ is the forbidden spacetime dimension. The
nearest possible $D$,  $D=5$, is of special interest: in this case
minor $H^{-1} H^a{}_\mu$ simply coincides with $h^a{}_\mu$; that
is, a contra-singularity simultaneously implies a co-singularity
(of high corank), but that is impossible! The possible
interpretation of this observation is:
 for the unique equation,
 contra-singularities are impossible if $D\,{=}\,5$
 (perhaps due to some specifics of
\emph{Diff}-orbits on the $H_a{}^\mu$-space).
  This leaves no room for any changes in the theory (assuming
 Nature abhors singularities).

 Linearization of eq-n~(\ref{estar}) looks like a  $D$-fold Maxwell
 equation, where infinitesimal diffeomorphisms serve as the set of
 gauge transformations; so, the number of polarisations is
  $D(D\,{-}\,2)$.

  And the unique equation  (like any other
  compatible system with the same structure of identities,
  {\em i.e.}, the
  same dimensionalities of the symbols, $\dim G_{2+r}$,
  see \cite{pommaret, on}) has 15 polarizations with $D\,{=}\,5$;
  it turns out that they are separated into five
  subclasses of quite different features.

\bigskip\noindent
{\bfseries\S3.\hspace{0.7em}Tensor $T_{\mu\nu}$ and post-Newtonian
effects (Pauli's questions to AP).}\quad
 One might rearrange ${\bf E}_{(\mu\nu)}{=}\,0$ picking out
 (into the LHS) the Einstein tensor,
 but the rest terms are not
 a proper energy-momentum tensor: they contain linear terms
 $\Phi_{(\mu;\nu)}$ [no positive energy (!); instead one more
 presentation of `Maxwell equation' (\ref{max}) is
  possible --- as the covariant divergence
  of a symmetrical tensor]:
 \be\label{sym} {\bf E}_{(\mu\nu)}+ 2g_{\mu\nu}{\bf E}_{\l\l}=
 -G_{\mu\nu}-\fr23\Phi_{(\mu;\nu)}+(\L^2\mbox{-terms})=0.
  \ee
 However, the prolonged equation
${\bf E}_{(\mu\nu);\l;\l}$ can be written as $RG$-gravity:
 \be \label{tmunu}{
 G_{\mu \nu
;\lambda ;\lambda }+ G_{\epsilon \tau} (2R_{\epsilon \mu \tau \nu
} - \fr12g_{\mu \nu }R_{\epsilon \tau }) =T_{\mu\nu} (\Lambda
'^{2},\cdots), \ T_{\mu\nu;\nu}=0; }\ee
 up to
quadratic terms,  \ ${ T_{\mu\nu}\,{=}\,
\fr29(\fr14g_{\mu\nu}f^2\,{-}\,f_{\mu\l}f_{\nu\l})\,
 {+}\,B_{\mu\eps\nu\tau}(\L^2){}_{,\eps\tau}
 } $\, \cite{on};
tensor $B$ has symmetries of the Riemann tensor, so the term $B''$
adds nothing to the $D$-momentum and angular momentum.

This equation (\ref{tmunu}) follows also from the least action
principle. The Lagrangian is quadratic in the field equations,
 {\em i.e.}
is trivial [one should use the trace eq-n, (\ref{trace}), and the
identity  $R\,{=}\,{-}2\Phi_{\mu;\mu}{+}\, (\L^2)$; $D\,{=}\,5$]:
\[L=  {\bf E}_{(\mu\nu)}^2 -7{\bf E}_{\l\l}^2
 \equiv \]
 \be \label{lagt}
 \equiv  R_{\mu\nu}G^{\mu\nu} + \fr19 f_{\mu\nu}^2 +
 \fr49[(3G_{\mu\nu}-\Phi_{\mu;\nu})
  \Phi_{\mu}+\Phi_{\l;\l}\Phi_\nu]_{;\nu}+
 (\L'\L^2,\, \L^4). \ee
After exclusion of covariant divergences, the main, quadratic
terms look like a modified gravity (higher terms can add only a
trivial quadratic contribution, like $B''$, to the energy-momentum
tensor).

 This Lagrangian is trivial, as well as all its Noether currents;
  it also means that the  contribution of gravitation
  to the `total energy' is
  negative and the `total energy' is strictly zero.
  (All this Lagrangian issue follows as a mere bonus, without any
  {\em ad hoc\/} activity\,!)

It is worth noting that:
\\
(a) the theory does not match GR, but reveals $RG$-gravity
 [of course,
eq-n~(\ref{tmunu}) does not contain all the theory];
\\
(b) only $f$-covariant (three transverse polarizations in
$D\,{=}\,5$) carries $D$-momentum and angular momentum
 ({\em ponderable} or {\em tangible} waves);
  other 12 polarizations are {\em imponderable}, or
{\em intangible}. This is a very strange thing, impossible
 in the Lagrangian tradition; how to quantize\,?
 \\
(c) $f$-component feels only the metric and $S$-field,
 see (\ref{max}), but $S$
has effect only on polarization (`spin') of $f$: $S_{[\mu\nu\l]}$
does not enter the eikonal equation, and $f$-waves moves along the
usual Riemannian geodesics;
 \\
(d) it should be stressed that the $f$-component is not the usual
(quantum) EM-field, but just an important covariant responsible
for energy-momentum (there is no gradient invariance for $f$
\cite{on}).

 Another strange feature is the linear instability
 of the trivial solution:
 some {\em intangible} polarizations grow linearly with time
 in the presence of
 {\em tangible} $f$-waves. Really,  the linearized
 eq-n~(\ref{ue}) and identity (\ref{iden}) yield
 (the following equations should be understood as linearized):
 \[
 \Phi_{a,a}=0, \  
 3\Lambda_{abd,d}= \Phi_{a,b}-2\Phi_{b,a},
 \ \Lambda_{a[bc,d],d}\equiv0,\]
 \be \label{inst} \Rightarrow \
 \Lambda_{abc,dd}=-\fr23 f_{bc,a}\, . \ee
 The last D`Alembert equation has the \emph{source} in its RHS.
  Some components of $\Lambda$
 (most symmetrical irreducible parts,
 as well as the Riemann curvature)
 do not grow because
 (linearized equations again)
 \[{ S_{abc,dd}=0, \ \Phi_{a,dd}=0, \ \,
   f_{ab,dd}=0, \ R_{abcd,ee}=0. }\]
  However the least symmetrical $\L$-components
  (triangle Young diagram), in fact only three polarizations
  of them
  which are to be called $\L^{\!\bullet}$-waves
  (three growing but intangible
   polarizations), do go up with time
  if  the ponderable waves (three $f$-polarizations)
  do not vanish. This should be the case for
  solutions of general position. Again, {\em nothing},
  {\em i.e.}\/ almost empty space, is practically impossible here\,!
  On the contrary, the three growing polarizations,
  $\L^{\!\bullet}$-waves,
  should result in strong nonlinear effects, and it is of special
  interest if some space regions can witness more instability,
  more nonlinearities, in comparison with other regions.

  The forth polarization of vector $\Phi_\mu$ [the fifth
  one is `eaten' by the trace equation, (\ref{trace})] is the
  (only) longitudinal polarization relating to the gradient
  part: $\Phi_{\mu}\,{=}\,\Psi_{,\mu}$. One can formally write the
  evolution equation for the longitudinal polarization
   [see eq-n~(\ref{trace})],
  \[ \square \Psi = \L^{\!\bullet}{}^2+\cdots\, ,\]
  indicating that the strong $\L^{\!\bullet}$-polarizations squared
  do contribute to
  the source for the longitudinal polarization.

  [Interestingly, the linearized equations (\ref{ue}) loose its
  trace part if $D\,{=}\,4$ (forbidden dimension; still one can add
  eq-n $\Phi_{\mu,\mu}\,{=}\,0$ `by hand') and in this case there
  is a new symmetry --- with respect to infinitesimal conform
  transformations which serve as the gradient transformations of
  field $\Phi_\mu$, and, therefore, eliminate the longitudinal
  polarization, so to say.]

  The skew-symmetric tensor $S$ is responsible for three
  polarizations. One can introduce the pseudo-tensor
   \[ \tilde{f}_{ab}=\fr16\,\ve_{abcde}S_{cde}; \]
 then, from eq-n (\ref{skew}) and the totally skew-symmetric part
 of identity (\ref{iden}), it follows (again a Maxwell-like system):
 \[ \tilde{f}_{[\mu\nu;\lambda]}=0\,, \ \,
   \tilde{f}^{\mu\nu}{}_{;\nu}=\fr18\,h^{-1}\,
   \ve^{\mu\nu\l\ve\tau}\L_{a\nu\l}\L_{a\ve\tau}\,.\]
 So, we have just three $S$-polarizations.

 Three $\L^{\!\bullet}$-polarizations correspond to tensor $\L$ of
 a specific, gradient (or rotor-gradient) form:
 \(\, \L_{\ve\mu\nu}\,{=}\,A_{[\mu,\nu];\ve}\,.\)

 At last, there remain five polarizations; this is just equal to the
 number of gravitational (Weyl) polarizations (in 5$D$);
 the evolution of
 these waves, see eq-ns (\ref{riem}) and (\ref{sym}), again has
 $\L^{\!\bullet}{}^2$-terms in its RHS (as a source) --- this time organized
 as a tensor relating to the square Young diagram (symmetry of the
 Weyl tensor).

 The current for the $f$-waves is just $Sf$-term, see (\ref{max}),
 therefore these waves are most weak, that is, their amplitude,
 $a_f$,
 should be smaller than the amplitudes of all other polarizations,
  \[ a_{\L^{\!\bullet}}\gg a_{W}, a_S, a_L \gg a_f\, \
 \, (3_{\L^{\!\bullet}}+5_W+3_S+1_L+3_f=15) . \]
 If some form of reduction to 4$D$ picture takes place, there
 could come forth  eight `preferable' polarizations:
 $2_{\L^{\!\bullet}}+2_W+1_S+1_L+2_f=8$.

\bigskip\noindent
{\bfseries\S4.\hspace{0.7em}Expanding
 O$_4$-symmetrical solutions and cosmology.}\quad
  The great symmetry of AP equations gives scope for symmetrical
 solutions. In contrast to GR,  eq-n~(\ref{ue}) has
non-stationary spherically
 symmetric solutions ($L$-waves).
An $O_4$-symmetric field  can be generally written \cite{on} as
\begin{equation}  \label{spsy}{
 h^{a}{}_{\mu }(t,x^i)=
  \left(%
\begin{array}{cc}
  a & b\,n_i \\
  c\,n_i & e\,n_i n_j+ d\Delta _{ij}\\
\end{array}%
\right); \  \ n_i=\frac{x^i}{r}; }
\end{equation}
 here $i,j\,{=}\,(1,2,3,4)$,
  $a,\ldots,e$ are functions of time, $t=x^0$, and radius
 $r$, $\Delta_{ij}\,{=}\,\delta_{ij}\,{-}\,n_i n_j, \
 r^2\,{=}\,x^i x^i$.

 As functions of radius, $b,c$ are odd
  while the others even;
  the boundary conditions are: $e\,{=}\,d$ at $r\,{=}\,0$,
 and $h^a{}_\mu\,{\to}\, \delta^{\,a}_\mu$ as
  $r\,{\to}\, \infty$.
Placing in (\ref{spsy}) $b\,{=}\,0, e\,{=}\,d$ (another
interesting choice is $b\,{=}\,c\,{=}\,0$)
 and making integrations one arrive to the next system
 (resembling Chaplygin gas dynamics; dot and prime
 denote derivation on time and radius, resp.)
\begin{equation}\label{gas}
A^{\cdot}=AB^\prime -BA^\prime +\fr{\,3}r AB, \  B^{\,\cdot
}=AA^\prime -BB^\prime -\fr{\,2}r B^{2},
\end{equation}
where $A\,{=}\,
 a/e\,{=}\,e^{1/2},\ B\,{=}\,{-}\, c/e\, $.
This system has non-stationary solutions, and a single-wave
solution (of `proper sign') might serve as a suitable (stable)
cosmological expanding background. 
  The condition $f_{\mu\nu}\,{=}\,0$ is a must for
   solutions with such
 a high symmetry (as well as
 $S_{\mu\nu\l}\,{=}\,0$); so, these $O_4$-solutions
 carry no energy, weight nothing ---
 some lack of \emph{gravity}\,!

 A more realistic cosmological model might look like a single
 $O_4$-wave
 (or a sequence of such waves) moving along the radius and being
 filled with a sort of chaos, or an ensemble of stochastic waves,
 both tangible (\emph{weak}, $\Delta a_f\,{\ll}\,1$) and
  intangible ($\Delta a_{\L^{\!\bullet}}\,{<}\,1$,
  but intense enough that  to give non-linear
  fluctuations with $\Delta h\,{\sim}\,1$).
  Development and examination of stability
 of this model would be an interesting problem.
 The metric inhomogeneity in a giant $O_4$-wave
 can serve as
 a time-dependent `shallow dielectric guide' for that weak
 $f$-waves. The ponderable  waves (which slow down
the large wave a bit) should have wave-vectors almost tangent to
the $S^3$-sphere of the wave-front  to be trapped inside this
shallow wave-guide; the imponderable waves can grow up, and partly
escape from the wave-guide,
 and their wave-vectors
can be less tangent to the $S^3$-sphere.

The waveguide thickness can be small for an `observer' in the
center of $O_4$-symmetry, but in co-moving coordinates it can be
very large, but still smaller than the current radius of the
$S^3$-sphere,
 $L\,{\ll}\,R$.

 This picture leads to the anti-Milne cosmological model
  \[a\,{=}\,a_0(1\,{+}\,H_0\,t)\, , \
  \ k\,{=}\,{+}1\, , \]
  which
 can describe well the SNe Ia redshift  data \cite{sn}.

\bigskip\noindent
{\bfseries\S5.\hspace{0.7em}Conclusion.}\quad
  One could conclude that the proposed theory is really beautiful
 and simple (like the chess rules), but its solutions are very
 complex (like chess games) and inevitably non-linear; the usual chess
 wisdom ``the less pieces the easier to play'' is not actual here
 because of the linear instability of the trivial solution.
 The great symmetry of the field equations and the fact that the
 frame field components belong to a single irreducible
 representation, as well as the unique possibility to be safe from
 singularities emerging in solutions, all these make the theory
 very attractive (the really single field theory).

  We have seen that both the extra dimension and
  the `new degrees of freedom',
 those `non-gravitational' polarizations, are not so terrible and
 can be incorporated into an interesting and promising picture.
 (It is possible that the `new degrees' does not influence the
 motion of `heavy bodies', or the ponderable waves, which still
 should move along the usual Riemannian geodesics.)

 The existence of a covariantly conserving stress-energy tensor
 (with a positive energy) is of great importance; this leads to
 (at least approximate) conservation laws, when a solution has
 (approximate, large scale)
 symmetries and Killing vectors.

  The presence of growing polarizations  does not mean the time
irreversibility --- their amplitudes
 (spatial Fourier harmonics) grow in both time directions
(like the amplitude of a frictionless oscillator under action of a
resonance force). But on the background of the cosmological
expanding $L$-wave [longitudinal wave which breaks the time
reversibility; symmetry of solutions is always (spontaneously?)\
smaller than the symmetry of equations] this instability yields a
kind of arrow of time. In other words, this is a sort of a
mechanism of transforming a continuous form of information
(inscribed in tiny details of weak vibrations of the frame field)
into a more non-linear form, and ultimately into a discrete form
of information carried by topological quanta, because, as it turns
out \cite{tc,on}, nonlinear field configuration can carry discrete
information --- topological charges and quasi-charges.

The topological issues in this theory are especially interesting
but require more efforts for exploration and much time for an
exposition.
 The tentative
phenomenology of topological quanta (a kind of `topological
 Brownian particles', which are non-local entities due to the huge
 extra dimension, and  can really move, their different parts,
  along different 3d-paths
 simultaneously) on the stochastic cosmological background could
 have a derivative Lagrangian with
  phenomenological (quantised or quantum) 4d-fields accounting somehow for
the contribution of quanta to the Lagrangian (\ref{lagt}).
 One might hope that this phenomenology can look  simple like a
 quantum field theory
[for with much chaos comes much simplicity; the superposition
principle comes as a result of summation along the extra
dimension, due to the almost tangent motion ({\em i.e.}
perpendicular to the extra-dimension) of tangible $f$-waves].

 Some qualitative conclusions and even predictions are already
 possible \cite{on,tc}, like the absence of spin zero
 elementary particles, as  well as supersymmetry.

It should be stressed that the theory itself has no `fundamental
parameters' or `constants', and all phenomenological `constants',
or very slowly (logarithmically?) varying `global parameters'
should emerge as  almost constant global parameters of a solution
itself.
 [The presence of dimensionless small parameters, like $a_f^2$,
 can potentially shift significantly the usual interpretation of
 the gravitational constant as the Plank scale (squared).]

 The recommended theory is very different from General Relativity, it
 has the extra-dimension and leads to a modified gravity;
 nevertheless, simple estimations \cite{nlaw}, with simple assumptions about
 the matter distribution along the extra dimension,
  show that  the Newton's law, $1/r^2$, can be reproduced at small distances,
 while the other law, $1/r$, is valid for large distances,
 $r\,{>}\,L$.
 The process of gravitational waves generation in this theory is also very
 different from the prescriptions of the General Relativity
 theory, especially for `short waves', $\lambda\,{<}\,L$.

\end{document}